# Political sense-making in “individual” human-technology-world relations: the case of LLM-generated autobiographies

**Anthony Longo and Olya Kudina**

**Reference:**

**Abstract**

As AI systems increasingly mediate how people describe, recognize, and present themselves, their politics cannot be reduced to what remains hidden inside the “black box.” This chapter argues that AI’s political significance emerges within technologically, intersubjectively, and temporally mediated interpretive practices. Bringing Arendt’s political philosophy into conversation with Gadamerian hermeneutics, the chapter develops a postphenomenological relational-temporal account of political sense-making across Human-Technology-World relations. Against the charge that postphenomenology remains individualistic and using the case of LLM-generated biographies, the chapter shows how seemingly private human-LLM encounters are worldly and collective. When users ask systems such as ChatGPT to describe them, they do not simply receive algorithmic representations. They interpret, contest, edit, share, and publicly test these descriptions. In doing so, people move interpretively among affective resonance, friction, public re-articulation, and the sedimentation of meaning. This means that the so-called “micro-level” of postphenomenological analysis is already saturated with publicity: affective uptake becomes communicable, reflection becomes accountable, and uses sediment into shared standards of appearance. Simultaneously, we show that once mediation is treated as relational and temporal, the political is not an add-on at the “macro” level; it is intrinsic to the very movement through which technologies amplify and reduce, stabilize and unsettle, and thereby compose a shared world in which judgment can take place. The chapter finishes by discussing several design and policy implications for AI of such political foregrounding in postphenomenology.

## Introduction

Recent postphenomenological engagements with Artificial Intelligence (AI) have significantly expanded the scope of mediation theory, with scholars exploring how algorithmic systems participate in shaping human experience, interpretation, and moral orientation (e.g., Benjamin et al. 2021; Hongladarom 2020; Yaghoobian 2025). Building on the classical insight that technologies are constitutive of perception, this literature investigates AI’s distinctive modes of mediation through its opacity, adaptability, and capacity for interpretive agency. Across these studies, the familiar relational H-T-W schemata of postphenomenology, denoting multiple possibilities for Human-Technology-World relations, are being reconfigured to address how machine learning, facial recognition, and Large Language Models (LLMs) alter the dynamics of intentionality and sense-making. Philosophers have described AI as a hermeneutic partner (Hongladarom 2020), as enabling “composite intentionality” between human and machine (Matei 2024; Verbeek 2008), or as operating through active user interfaces that afford quasi-embodied relations (Gerlek and Weydner-Volkmann 2025). Postphenomenology captures these AI-inflicted conceptualizations through the encompassing dynamics of H-T-W sense-making, whereby all technologies can be read through a hermeneutic and world-disclosing lens: technologies do not merely mediate access to reality but also always interpret and generate meaning (Kudina 2021; 2023). The manifold developments in AI invite augmenting this dynamic and relational sense-making further and applying it to the case of AI specifically. Through algorithmic operations of filtering, pattern recognition, and prediction, AI constructs what Hongladarom (2020) calls a "double mediation," in which the human perceives a world already interpreted by the machine. This renders AI a co-producer of sense, or what Hongladarom and van der Vaeren (2024) describe as a form of post-anthropocentric hermeneutics, where interpretation becomes a distributed and performative process. Together, these analyses propose a wide range of accounts of how AI mediates both experience and meaning. Yet, a persistent question remains underexplored: what kind of *shared* world comes into being through these mediations, and what *political* order of visibility, agency, and judgment do they sustain?

The question of world-disclosure touches on a long-standing debate concerning postphenomenology's alleged political deficit. Some have argued that while postphenomenology may excel at analyzing concrete human–technology relations, it remains largely blind to the structural, collective, and historical dimensions of technological mediation (Bantwal Rao et al. 2015; Kaplan 2009; Feenberg 2020; Coeckelbergh 2022). From this perspective, its methodological focus on micro-level experience allegedly fails to account for how technologies shape public life, power, and collective agency. Some have, therefore, called for expanding the framework from "human–technology" to "human*s*–technology" relations, to capture the formation of social and political collectives through technological design (Feenberg 2020; Dameski et al. 2024). Similarly, Kaplan (2009) cautions that without engaging political economy and historical critique, postphenomenology risks remaining "naïve to power." Yet, a growing group of scholars has sought to address this limitation by connecting postphenomenology's basic assumptions with relational approaches to politics and power. Verbeek (2013; 2020), for instance, has reinterpreted postphenomenology through a Foucauldian lens to conceive of power as a constitutive feature of mediation rather than an external force. Claassen (2024) goes back to phenomenological roots of the tradition and, with the help of Schutz, showcases the primacy of social situatedness of “the human“ in H-T-W scheme, how it can never be an isolated “I.” A distinct effort to showcase politics within postphenomenology focused on bridging  Arendt’s political phenomenology with postphenomenology to situate technologies within processes of public appearance, intersubjectivity, and collective sense-making (Baş 2026; Longo 2024; Verbeek 2020). From this perspective, mediation is never apolitical but the very terrain on which political subjectivity and worldhood are negotiated.

Much of the critique directed at postphenomenology’s alleged political deficit rests on an overly static image of mediation. Yet this reading overlooks how several contemporary developments within postphenomenology have already temporalized the human–technology relation. Some scholars have demonstrated that mediation is not a fixed structure but a dynamic process through which meaning and normativity evolve (Kudina 2021; Baş 2026; Longo 2025). Rather than separating private experience from public life, a “political postphenomenology” (Verbeek 2020) traces how technological mediation carries experience across these domains—how affective, embodied, and moral orientations sediment into shared practices, discourses, and forms of collective visibility. From this perspective, the political dimension of technology does not stand outside postphenomenological analysis but already sits in the temporal movement of mediation itself: in the ways technologies modulate the transitions between immediacy and reflection, between lived intimacy and public articulation. By taking temporality seriously, postphenomenology can thus articulate how the political emerges from within the ongoing transformations of mediated experience, rather than being imposed from the outside.

This temporal and relational view of mediation also offers a critical rejoinder to prevailing accounts of algorithmic subjectivity that frame AI as displacing or hollowing out human self-understanding. Here, algorithmic systems are often conceived as fundamentally “replacing” the human faculties of reflection, judgment, and self-discovery. Fisher (2022) argues that algorithmic knowledge seeks to “bypass” subjectivity, competing with and ultimately undermining our capacity for critical self-relation. Similarly, Joseph (2025) describes this as a “collapse of introspection”, in which the “inner work” of self-understanding is delegated to algorithmic systems. Eisikovits and Feldman (2022) warn that such replacements of everyday judgment by AI threaten to erode our ability to deliberate, while Gertz (2018: 55) characterizes this relation as one of nihilism, where users relate to algorithmic media “as if they were our destiny.” Becoming a subject no longer seems to be a human endeavor but is displaced by the AI-driven production of “inferential clusters–pulsing patterns and propensities distilled from distributed data points” (Petersmann and Van Den Meerssche 2023: 121), formatting us into “statistical individuals” (Bjerring and Busch 2024). Across these perspectives, users are conceptually constructed as largely passive victims, subjected to automated forms of interpretation that replace or pre-empt human meaning-making.

Against this backdrop, our analysis both challenges the assumption that algorithmic mediation necessarily entails such a displacement of selfhood and showcases the inevitability of the political dimension in postphenomenology. Taking the case of users generating, adopting, and testing autobiographical narratives with large language models (LLMs), we show that meaning is not simply transferred from machine to human (or *vice versa*) but unfolds across a recursive loop in which the human, technological, and worldly poles co-evolve. Rather than collapsing introspection and subjectivity altogether, these relations reveal how reflection and politics emerge *through* the temporal movement of mediation itself— where affective resonance, friction, public articulation, and sedimentation become movements in the ongoing constitution of meaning. In this sense, autobiographical sense-making through AI offers a microcosm for exploring the *technopolitical mediation* (Baş 2026) of selfhood: how private reflection becomes a public appearance mediated by algorithms. We start off by presenting a critique of the "black box" metaphor of AI as an entry point, since it epitomizes the assumption that AI acts as an opaque external force replacing political sense-making. Against the widespread calls for "opening up" the black box to allow public and political scrutiny, the next section argues that political sense-making does not wait for perfect transparency; it already takes shape in interpretive practices. After that, we unpack and counter the key assumptions underpinning the "black box" metaphor to develop an alternative hermeneutic approach to political sense-making by connecting Arendt's political hermeneutics to postphenomenology. With this approach, we then provide a more nuanced and relational analysis of autobiographical sense-making with LLMs, and conclude with a brief reflection on the political payoff of postphenomenological AI research.

## 1. Black-boxing the political

Unlike many objects of study for social scientists and humanities scholars, algorithms are not immediately readily available for scrutiny. Machine Learning (ML) algorithms, at work in contemporary AI, operate *through opacity* (McQuillan 2018: 2). This is because, to make pattern recognition within a given dataset possible, an algorithm must "understand" (i.e., computationally process) all the relations between different input variables within that dataset to detect statistical correlations and distributions. The dimensionality of a neural network refers to the number of possible combinations among different variables. This makes the pattern recognition that ML algorithms perform dependent on the computation of variables in dimensions that surpass human intuition (Benjamin et al. 2021: 2). In other words, the opacity of algorithmic decision-making is not a "flaw" of the system but precisely a condition for generating insights beyond human capacity. Although its high dimensionality may seem like a *curse* for human users and designers, an increase in the combined dimensionality of neural networks is actually what powers the processing capacities of algorithms (Gorban and Tyukin 2018: 2).

In response to the complexity and inaccessibility of ML operations, scholars have described the algorithms curating and organizing information as “black boxes” (Dobson 2023: 433; Leerssen 2020: 2). The different uses of this metaphor correspond to different kinds of claims being made about algorithms. In the first instance, the metaphor re-renders algorithms from a purely technical-abstract set of instructions into an “epistemological problem” (Bucher 2018: 43): “As outsiders, “we” do not know for sure what is happening behind the curtain” (Seaver 2019: 417). Pasquale (2015) popularized the concept of the “black box,” originally highlighting a critical issue of access and understanding for both researchers and users: “we can observe its inputs and outputs, but we cannot tell how one becomes the other” (Pasquale 2015: 3). This concern, however, predates Pasquale’s work. As early as 1974, McCarthy identified the opacity of AI systems as its fundamental “disease” (McCarthy 1974). As several scholars have indicated, the black box problem is not a single easily solved feature of algorithms but includes a range of distinct but interconnected issues, like opacity, unpredictability, justification problems, lack of technical literacy, and the sheer size of algorithmic calculations (see Brożek et al. 2024; Burrell 2016; Christin 2020).

In this regard, the metaphor of the “black box” has also emerged as a central source of normative critique, positing algorithms as a *political problem*. This approach emphasizes the challenges algorithmic opacity poses for academic inquiry, accountability, and regulation (Eubanks 2019; O’Neil 2016; Zuboff 1988). Calls for “algorithmic transparency” frequently mobilize the black box metaphor, highlighting the inaccessibility of source code and decision-making processes as fundamental barriers to scrutiny and accountability (Kitchin 2017: 20;

Leerssen 2020: 2). Efforts to "open up" the black box (e.g., Diakopoulos 2015) rely on the assumption that making algorithms visible or public is a prerequisite for meaningful ethical and political interaction with them. This perspective is echoed in the field of AI scholarship, where the black box problem is often reframed under the label of "explainability" (Guidotti et al. 2018; Miller 2019; Vilone and Longo 2021), leading to the development of the Explainable AI (XAI) movement (Gunning and Aha 2019). The XAI movement underscores the notion that algorithms performing critical tasks should become "white boxes" (Brozek et al 2024: 429) or "glass boxes" (Rai 2020), transparent not only to their creators but also to end-users and those affected by algorithmic decisions. These approaches rest on the assumption that gaining access to the "true" algorithm concealed within the black box is essential for enabling meaningful political and ethical scrutiny (Seaver 2019: 415; Matzner 2023: 11). The opaque nature of algorithms is seen as a fundamental barrier, preventing the public from engaging in "meaningful and active interaction with the hidden algorithms that regulate its behavior" (Perel and Elkin-Koren 2017: 181).

The "black box" metaphor thus performs double duty: as an *epistemic* diagnosis (i.e., we do not know how it works) and as a *political* critique (i.e., opacity blocks public scrutiny). Together, they construct "the algorithm" as an *ontological* problem. Such framings are never neutral; they shape algorithms as particular objects of study (Lakoff and Johnson 2011; Verrycken et al. 2013). Seaver (2017: 5) notes that different starting perspectives shape different ways to study and enact an algorithm, e.g., through mathematical or ethnographic analysis. The black box metaphor enacts algorithms as self-contained, opaque "things-in-themselves," independent from human perception and intervention. By positing algorithms as enigmas to be unboxed, it risks reifying AI and detaching it from the human and institutional contexts of development and deployment, directing efforts toward uncovering a "hidden core [with] secret, embedded values within" (Gillespie 2016a: 52).

Beneath the epistemological and political dimensions of the black box metaphor lie two philosophical premises that quietly determine how the relation between technology and politics is understood. (P1) First, the black box discourse admits to an *interiorization* of politics: what is politically salient is presumed to hide *inside* the artefact (e.g., its code, parameters, model, training data). Therefore, *access* to this "interior" of the algorithmic infrastructure (through transparency, explainability, audits) is cast as the primary political remedy. (P2) Second, it commits to an *exteriorization* of the algorithm: it treats "the algorithm" as a bounded, discrete, and autonomous agent-like thing that intervenes from *outside* human experience and agency. Together, they conceptually construct a picture of user-algorithm relations as inherently asymmetrical and divided: (P1) political meaning is hidden inside the algorithm, (P2) users cannot access the algorithmic interior, (C) therefore, users are fundamentally politically alienated. In methodological terms, the *political* status of algorithms (i.e., what role they play in the political realm) is inferred from their *ontological* status (i.e., as "black box").

We claim that the so-called political problem of the black box arises from an artificial and forced conceptual construction of the relationship between technology and politics. It is forced because it conflates the ontological registers of technical transparency (i.e., "how it works") with political meaning (i.e., "what is at stake"). Ryle conceptualized such conflations as a "category mistake": to mistakenly "allocate those concepts to logical types to which they do not belong" (Ryle [1949] 2009, 7). If, for example, a visitor to Oxford University asks to be shown "the university" after touring the libraries, colleges, auditoria, etc., they treat an institutional complex as though it were one more physical building. This request is not merely mistaken, it is, according to Ryle, *unintelligible*: it mislocates the object of inquiry in the wrong logical space. The black box metaphor commits a similar mistake when it equates the "material disclosure" of the algorithm's source code and model with "political meaning." As we will argue in section 3, political meaning belongs to a different category than technical transparency. The two neither entail nor subsume each other. Code disclosure does not guarantee to deliver intelligibility if the released information is inscrutable to publics. Conversely, as we will argue in section 3, intelligibility may arise even under opaque conditions (e.g., when gig workers develop tactics for "gaming" ride-share surge pricing without ever seeing Uber's algorithm; cf. Bonini and Treré 2024: 162). To understand an algorithmic infrastructure, in a political sense, is to know how to evaluate, contest, and re-purpose its mediations, not necessarily to recite its interior mathematical structure. Rather than demanding to open up the black box, we should analyze how publics *already* encounter, interpret, judge, and react to algorithms before and beyond visibility and interaction with the code. The remainder of this chapter will connect Arendt and postphenomenology's theoretical anchors to establish this alternative approach, by rejecting respectively premise 1 (in section 2) and premise 2 (in section 3) and highlighting how politics is an inalienable part of postphenomenological human-technology relations.

## 2. Politics doesn't operate from the inside

In this section, we will analyze the category mistake identified in section 1 (i.e., the assumption that political meaning lies inside the algorithm, concealed behind its visible outputs) with the help of Arendt's political thought. This mistake can be traced back to a Platonic tradition of political thought. In Plato's metaphysics, the visible and changeable world of human affairs is understood as the realm of mere semblance, of shadows that conceal a higher, more stable reality of eternal forms or Ideas. Politics here becomes a "design issue primarily": if appearances are deceptive, the task of the political thinker is to look beyond them, to uncover the true form of politics and to *design* the public space accordingly. Through this Platonic move, political action, with all its uncertainty, spontaneity, and dependence on the presence and recognition of others, is replaced by the more controllable logic of making and governing according to an ideal model (Arendt 2018: 220ff). This substitution has far-reaching consequences. The political realm is no longer a space of shared appearance but a site to be shaped in correspondence with a truth that lies beyond it.

The same metaphysical logic underlies many contemporary debates on algorithms. Calls to "open up the black box" reproduce Plato's distrust of appearances by assuming that what we see on our screens is only a shadow of a hidden and truer reality. Digital processes are treated as *noumenal*—things that exist independently of perception and can only appear through the mediating interface (Vial 2019: 85). And when Meta's Ex-Chief AI Scientist LeCun insists that ChatGPT only *appears* intelligent while "in reality" doing nothing new, he echoes Plato's allegory of the cave: the enlightened expert who sees beyond the shadows to the true workings of the world (Coeckelbergh and Gunkel 2023: 2). Similar convictions continue to shape efforts of accounting for the politics of algorithms: by situating political meaning in an inaccessible "inside" realm of code. The result is a political theory that seeks control and certainty at the expense of the plural, unpredictable, and worldly character of political sense-making.

Against this Platonic distrust, Arendt proposes a radical rehabilitation of the sphere of appearance as the primary site of political sense-making. In her view, what appears to us is not a secondary or deceptive layer covering a truer reality, but the only domain in which human beings can disclose themselves, encounter one another, and make sense of the world they share: "Nothing and nobody exists in this world whose very being does not presuppose a spectator" (Arendt 1981: I.19). Arendt extends this insight into the political realm: what is at stake in politics is how the common world appears to a plurality of human beings. This plurality is expressed in the realm of opinion: "there are, or should be, as many different *logoi* as there are men, and all these *logoi* together form the human world" (Arendt 1990: 85). The task of politics, then, is not to pierce through appearances to reach an ultimate truth, but to sustain and orient ourselves within this plurality of appearances. Arendt thus redefines political meaning as something that arises from the interpretive activity of those who witness and judge events in the world: "the public realm is constituted by the critics and the spectators, not by the actors and the makers" (Arendt 1982: 63).

Arendt emphasizes that political sense-making (through spectatorship and judgment) is always mediated, both materially and intersubjectively. The "world" functions in Arendt's political theory as a twofold "in-between": first, as an objective in-between (*Dingwelt*) and, second, as an intersubjective in-between (*Mitwelt*). From the material side, political meaning always depends on the work of *homo faber*, the maker who transforms fleeting acts into durable forms that can appear to a plurality of human beings. Political action is inherently ephemeral—it "comes and goes, leaving no trace", unless stabilized by artefacts, records, and stories (Arendt 2018: 95). She writes: "the whole factual world of human affairs depends for its reality and continued existence upon the presence of others who have seen and heard and will remember, and upon the transformation of the intangible into the tangibility of things" (Arendt 2018: 95). Yet such reification is never sufficient in itself: artefacts and records only become meaningful when they are interpreted and judged by spectators. The work of *homo faber* thus prepares the material conditions for remembrance, but it is the act of spectatorship that reanimates these traces within a shared horizon of meaning.

If the *Dingwelt* gives fleeting appearances a sense of endurance, the *Mitwelt* (the world of others) makes it communicable. Arendt situates judgment within the basic condition of plurality: we never think or evaluate worldly matters in isolation but in relation to others" viewpoints. Acting and judging are "decentered" activities, whose meaning exceeds the control of their author (Tchir 2017: 70). This non-sovereignty is not a limitation but precisely the starting point for political judgment: judgment responds to something in the world that elicits our attention—an event, an image, an utterance—that first moves us affectively and only later becomes available for reflection (Arendt 1982: 68–69; cf. Borren 2010: 23). In this movement, we appeal to what she calls *sensus communis* (or "common sense"): not an explicit or empirically available collection of others" opinions, but the imagined standpoint of others to whom one seeks to make one's view intelligible, as acquired through experience in the respective sociocultural context (Arendt 1982: 70-72). By taking into account how something might appear to others, we "fit ourselves into a community" that is both presupposed and enacted through judgment (Borren 2013: 248). In this sense, the public realm is not a battleground of competing opinions or a site for consensus-building but a "testing ground" (Arendt 1982: 40) in which we learn to orient ourselves among others and to render our experiences communicable. Together, these dimensions ensure that political meaning is co-shaped by the artefacts that endure and the others who interpret (cf. Baş 2026: 110ff).

This movement between private and public experience also involves a gradual shaping of who we are. For Arendt, acting, speaking, and judging in the public realm are ways in which people become visible as unique individuals (Arendt 2018: 178). The "who" only emerges in interaction, in how one appears to others. The disclosure of who we are, therefore, depends on the existence of a public space where people can appear before others. Loidolt (2018: 210–12) distinguishes three interrelated dimensions of this self-disclosure: (1) the pre-reflective experiential sense of selfhood that underlies all possible experience and perspective-taking; (2) the affective and stylistic way one finds oneself attuned to the world, the singular manner of inhabiting and enacting existence; and (3) the narrative and public articulation of selfhood through the stories one lives and that others tell. These layers are not sequential but intertwined: the pre-reflective and affective dimensions are articulated through narrative identity, while narrative identity, in turn, presupposes the worldly immediacy of lived experience. This Arendtian–phenomenological model replaces the fixed, interiorized subject of traditional political theory with a dynamic and worldly conception of personhood, i.e., one constituted through acts of appearance, spectatorship, and judgment in a shared world. It provides a crucial philosophical grounding for understanding how *political meaning*, along with selfhood itself, emerges *through* appearances. This emphasis on worldly appearance resists the assumption that "politics operates from the inside"; it instead reveals that political sense-making emerges in shared spaces of disclosure shaped by intersubjective and technological mediation alike. Establishing this is essential for examining the so-called "political deficit" in postphenomenology, which we will explore in section 3.

### 3. Algorithms don't operate from the outside

Arendt showed that political meaning depends on a double "in-between" that stabilizes fleeting appearances and makes them communicable to others. Yet, as Baş has argued, Arendt's account hesitates when it comes to contemporary *technologies* as material end-products of homo faber: she tends to associate modern technology with world-alienation, while attributing the stabilizing and inspiriting power of reification primarily to cultural artefacts, such as artworks and stories (2026: 110–18). Technologies, and how they help to shape specific subjectivities and worlds, is the exact focus of postphenomenology. Postphenomenology treats technologies as mediators that amplify and reduce certain perceptual aspects over others, exhibit multistability, and enter histories of appropriation and stabilization in concrete human-technology-world relations. If we take Arendt's account of the cultural products of homo faber as our starting point rather than her limited reading of "technology," we can see how artefacts stabilize the appearances of praxis *retroactively* and inspire further appearances *proactively* (Baş 2026: 118). Postphenomenology extends this Arendtian insight from cultural artefacts to material-technological ones (e.g., algorithms, interfaces, platforms), thereby giving us the mechanisms and temporal dynamics needed to track how political sense-making emerges and sediments through mediated loops of use.

The shared philosophical foundation in a *relational human-world ontology* allows us to connect Arendt’s political theory to postphenomenology (Loidolt 2018: 84–87; Verbeek 2011: 74). Postphenomenology radicalizes this insight by showing that technologies participate in this co-constitution, co-shaping how subjectivity and world arise in practice (Ihde 1990; Verbeek 2005). To uncover the dynamics of sense-making in this relational co-constitution and building on the hermeneutic scholarship of Gadamer, Kudina (2023) introduces the “hermeneutic lemniscate” as the principle of dynamic interpretation in the H-T-W relations (see Figure 1).

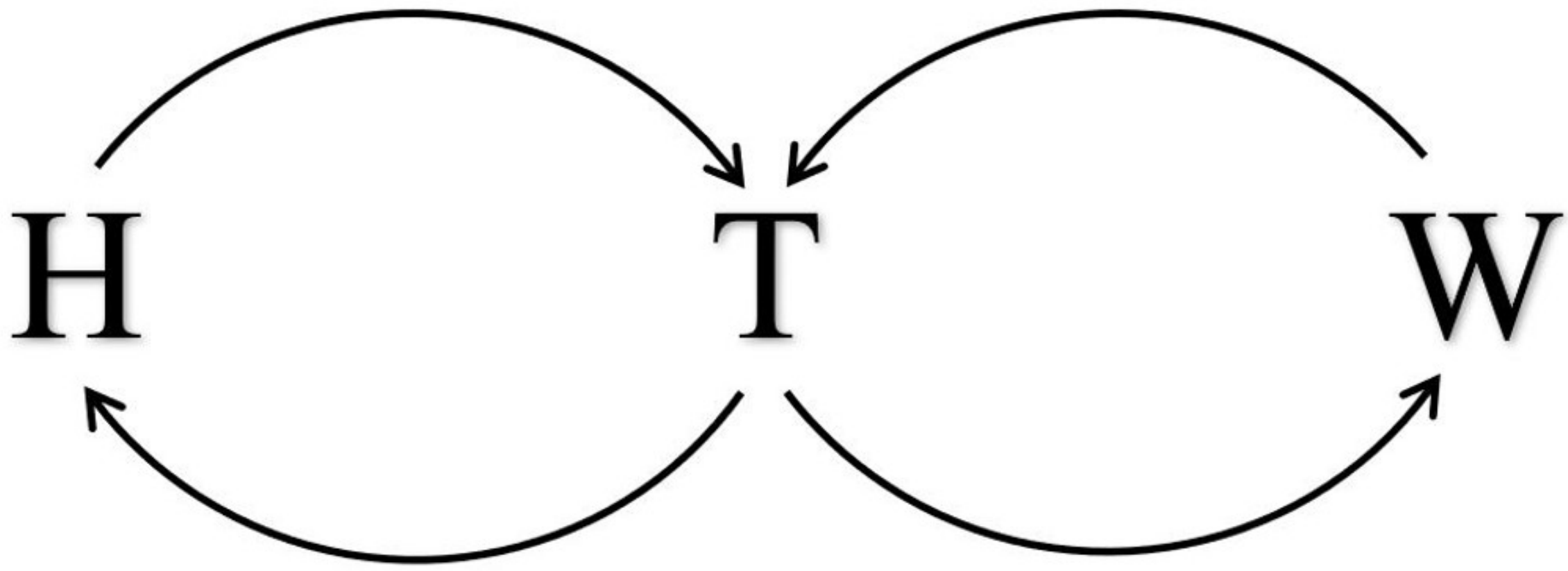

*Figure 1. Hermeneutic lemniscate principle of human-technology-world interpretation (Kudina 2023)*

This principle enriches the classic postphenomenological schema by highlighting the interpretive dimension in all types of H-T-W relations and the ongoing process of interpretation as ontologically relevant to the concerned human, technology, and the world insofar as they become (re)interpreted in the mutual encounter. This interpretation is continuous and only preliminarily stabilized to reflect the inherent dynamics and expanding horizons within each of its H-T-W parties. Such a continuous nature of sense-making due to the productive role of humans, technology, and the world in interpretation reflects

Gadamerian (and Heideggerian) temporal ontology, whereby every presently established meaning becomes a part of productive history to jumpstart future interpretations. Crucially, following Gadamer ([1975] 2004: 303), every person represents an interpretive horizon as a fusion of two dimensions: the sedimented past of one's own experiences and understandings (i.e., "our own past"), and the historical tradition that precedes us and shapes our shared mode of existence (i.e., "that other past"). The individual, thus, goes hand-in-hand with the collective.

For Kudina (2023), a productive way into the process of interpretation is by focusing on how people appropriate technologies: how they (re)interpret, integrate, and project meanings in relation to technologies. Such appropriation is both projective and practical: people form anticipatory relations to technologies before any direct or physical interaction with them, and these anticipations guide later interpretations in use. Importantly, this process co-evolves with a shifting understanding of oneself as self-understanding develops through engagement with the technological and social world, not prior to it. That is, we cannot avoid (re)inventing ourselves while making sense of technologies.

The hermeneutic lemniscate thus visualizes a dynamic loop in which technological mediation and self-interpretation unfold together, making explicit the temporal, reflexive, and transformative dimensions implicit in Arendt's notion of the space of appearance. The hermeneutic lemniscate makes clear that there is no "inside" or "outside" to technological mediation and that the loop is never closed due to the ever-expanding horizons of meaning that H-T-W interpretive parties represent. Meaning emerges *in between*—in the reciprocal adjustments and interpretations that take place throughout this ongoing relation. Because each pole co-constitutes the others, meanings remain open-ended and stabilize only provisionally until a new sociomaterial situation arises that questions the fitness of the previously established meaning. This is precisely what enables the "multistability" that characterizes all mediated experience. This hermeneutically substantiated postphenomenology fundamentally rejects the image of the algorithm (or any technology) as a self-contained "black box" acting upon an isolated user from the outside; instead, it situates political and experiential sense-making within the dynamic interplay of mediated relations.

However, algorithms are never neutral mediators. They reflect and reinforce pre-reflective user behavior, such as preferences, habits, and affective patterns, simultaneously reducing the plurality of perspectives and the time to reflect on this. Thus, algorithms are perceptually and politically salient. Longo (2025) shows how this dynamic can be traced through five recurring stages of user–algorithm engagement that map onto Kudina's hermeneutic trajectory through affective resonance, algorithmic reflection, disruption to sense-making, contestation, and production of folk theories. In affective resonance, users first encounter algorithms emotionally, and these moods and sensations orient further online participation. In algorithmic reflection, platforms mirror users" identities back to them, generating the experience of being "seen" by the system and giving rise to what Bhandari and Bimo (2022) call the algorithmized self—a sense of identity co-shaped by algorithmic feedback. Disruption occurs when algorithms over- or mis-amplify certain traits, providing friction or dissonance to user experience (e.g., "why is this on my feed?"), triggering reflective awareness of algorithmic mediation. Such moments often become sites of contestation, negotiation, and judgment, both individually and collectively, as users recalibrate their sense of agency and belonging within algorithmic publics (Couldry et al. 2016; Gillespie 2016b). In repeated loops of reflection and recalibration, users develop intuitive *folk theories* about algorithms (DeVito et al. 2017), which coalesce into broader "algorithmic imaginaries": shared frameworks for understanding how algorithms work and shape public life (Bucher 2018; Gran et al. 2021), similar to becoming a productive historical background for future interpretations in Kudina's analysis. These five stages exemplify the hermeneutic lemniscate in practice: rather than a one-way process of technological determination, user–algorithm relations form an evolving circuit of affective, interpretive, and normative attunement. Amplification and reduction also productively support political sense-making: they render both the self and the shared world intelligible through mediated acts of appearing and judging.

The political relevance of this interplay becomes clear once we recognize that algorithmic mediation structures not only what is seen but also what remains unseen, shaping the field of possible judgment. It is within this ambivalent structure of disclosure and withdrawal that we situate our case study of LLM-generated biographies in the next section to understand how processes of mediated self-discovery traverse the loop between private and public sense-making.

## 4. Autobiographical sense-making with LLMs

In the preceding sections, we have argued that the political meaning of algorithms does not reside in their internal workings but in the relations of sense-making through which people engage with them. Arendt's rehabilitation of appearance and spectatorship, together with the hermeneutically expanded account of technological mediation, provided the conceptual means to understand such relations as co-constitutive of both self and the world. This section applies that framework to a concrete case: the increasingly common practice of using LLMs, such as ChatGPT, to generate autobiographical descriptions. Here, LLMs produce narrative biographical accounts of users based on all their interactions and prompts, e.g., "describe me," or "tell me something about myself I don't know."

This case study allows us to examine how this act of human-technology interpretation entails the private–public movement through which experiences become communicable, testable, and thus real. In these encounters, users oscillate between private curiosity and public sharing, self-reflection and collective interpretation. The case exemplifies a new form of technological in-between, where selfhood is amplified and made visible to others. Rather than assuming a rivalry between two distinct epistemic agents (humans vs. algorithms) (e.g., Fisher 2022), our analysis shows how LLMs participate in the process through which the self can become publicly legible.

People increasingly use LLMs for generating biographies. In an illustrative blog post, Chandler (2025) asked ChatGPT to "describe me based on all our chats." The model generated a flattering yet strangely generic description—an "eloquent, reflective, intellectually curious" individual—with Chandler noting that the text "sounded like me without being about me." He concluded that ChatGPT "mirrors the tone of our exchanges more than my life," situating the AI as an interpersonal mirror rather than an epistemic knower (Vallor, 2024). Similarly, when Smith (2025) asked ChatGPT to "tell me something about myself I didn't know," she observed that it could only "connect dots I had already placed." Schwartz (2025) requested an image that would "describe his life." After correcting factual mistakes ("I'm not a photographer, I'm a writer"), the user remarked that the process felt like negotiating own portrayal.

These accounts already hint at a dialogical structure of meaning-making, inherent in the different ways people appropriate LLMs. When users treat AI as a confidant, an editor, or a mirror, they highlight the interpretative process that oscillates between the H-T-W axis. Even though the hermeneutic process starts from a user-LLM interaction, the interpretative act in parallel imbeds a constant exchange with the larger world, both through the ways LLMs imbed the collective information with all its prejudices, and hence, the constant presence of politics; and through the act of publicly sharing individual biographies, written by LLMs. By using LLMs to generate biographies, users appropriate these systems in ways that let them articulate what matters to them, what troubles them, and what they wish to contest in how LLMs represent them. In doing so, they also reflect on their identities and shape their self-understanding. This technologically mediated sense-making directly challenges the claim that postphenomenology is biased toward individual relations. What seems like an individual H-T-W relation has an inalienable collective dimension: interpretative acts with LLMs enrich the background of understanding we build for others and bring these representations into public reflection. The hermeneutic lemniscate expansion of postphenomenology only makes explicit what was already there: individual relations are of the common world, which is productively present in both individual use and the technology itself.

Furthermore, LLM biographical practices are echoed in communal settings. A Reddit thread titled "Asked ChatGPT to guess how people would describe me, and now I'm smiling like an idiot"[1] invited participants to share the model's characterizations. The resulting conversation combined irony and introspection: one user quipped that ChatGPT's description—"empathetic, thoughtful, occasionally self-critical"—was "basically everyone on Reddit," while another noted that the model "knows me better than my boss." The thread demonstrates how algorithmic self-writing moves quickly from individual, private confrontations to *collective* negotiation, producing what Gillespie (2016) calls a "calculated public." McAllister's (2025) op-ed "Can AI Write Your Memoir?" and a report on *The College Fix* (Kabbany 2025) on students outsourcing personal statements broaden this picture: they register a social normalization of algorithmic co-authorship in the presentation of self. Authorship and authenticity become shared and negotiable rather than private and proprietary.

Scholarly reflections document similar tensions. Jeremiah (2024) shows how writing with GPT-4 exposed affective patterns "of impatience, frustration, and [...] excitement," reshaping her self-awareness as a researcher. Santarpi (2024) analyzed how ChatGPT-based poetry sessions helped participants "recognize dormant feelings," while

1 https://www.reddit.com/r/ChatGPTPromptGenius/comments/1iowmn0/asked_chatgpt_to_guess_how_people_would_describe/ (last consulted 27 November 2025) risking "illusionary intimacy." Across these examples, users seek reflection and recognition, encounter uncanny intimacy and estrangement, and often re-share the results for public commentary. The "AI-authored self" thus becomes a hybrid artifact of sense-making—part self-portrait, part social performance—whose meaning crystallizes through the ongoing circulation in the H-T-W relations.

These collective negotiations can be read as digital enactments of Arendt's space of appearance. When users ask ChatGPT to "describe me," aspects of the private self become textual, shareable, and open to judgment; users also become spectators of their own appearance. Chandler's claim that ChatGPT "sounded like me without being about me" captures this split position, while Smith describes "reading my own portrait as if written by a friend who half remembers me." Fisher's (2022) worry that algorithms might "tell us who we are" treats identity as something to be known rather than as a dynamic appearance. Yet users' affirmations, corrections, irony, and public postings show that the "who" becomes legible only dialogically, echoing Arendt's idea that meaning stabilizes through exposure to plurality. Santarpi's poetry-therapy participants illustrate this shift: once discussed in group sessions, their ChatGPT-generated verses moved from private affective exploration to intersubjective significance. The biography, poem, or profile thus performs an Arendtian transformation from private feeling to communicable experience.

Having outlined the theoretical model of mediated political sense-making in sections 2 and 3, we can now trace how this model unfolds in the context of user–LLM encounters. When users prompt LLMs to "describe who I am" or to write a "personal narrative," they enter a recursive relation in which the human, technological, and worldly poles of experience are reconfigured through successive movements of affective appropriation, reflective correction, public articulation, sedimentation, and co-evolution, across which meaning emerges:

1. *Affective appropriation.* Initially, users approach LLMs with curiosity or a desire for recognition. Many describe the encounter as comforting or intimate, “like confiding in a non-judgmental other” (Jeremiah 2024). The model’s fluency and affirmation elicit what Bucher (2018) calls the “affective pulse of algorithms”: an immediate sense of being seen. Here, the human is shaped by affective uptake rather than reflective distance, while the technology functions as an embodied mediator, i.e., a surface of projection that remains largely transparent. The world appears as a space of possibility and affirmation, where users’ pre-reflective desires for recognition are technologically amplified. Hermeneutically, this produces a temporary fusion of horizons: self, technology, and world interlock in mutual validation. This confronts users with biases about oneself, the self-understandings they already inhabit, now mediated by LLMs and projected onto a world that appears to affirm them.

2. *Friction and reflective correction.* The interpretative loop shifts when initial resonance meets dissonance. A generated self-portrait may appear inaccurate, generic, or emotionally shallow; Chandler’s “eloquent but empty” reaction captures this friction. Surprise or disappointment signals a transition from passive absorption to active interpretation, as users correct, refine, and reinterpret the model’s output. Phenomenologically, the technology shifts from transparent medium to object of attention, moving from the background to the foreground of experience. The world disclosed is no longer a comforting mirror but a space of negotiation, disclosing the limits of algorithmic recognition. Users begin to read affirmations as selective and constrained, amplifying some traits while reducing complexity, difference, and conflict. This reduction compels users to clarify their self-understanding, transforming frustration into interpretive insight. In Ihde’s sense, the AI becomes a material hermeneutic device: through its misfires, users grasp aspects of themselves that had remained implicit.

3. *Re-articulation and public engagement.* In a further interpretative movement, users externalize reflection by sharing AI-generated self-descriptions online. Reddit threads and social media posts transform solitary interaction into collective interpretation. Comments such as "mine said the same thing—maybe we're all the same person" disclose a new phase: private dialogue becomes a plural scene of comparison, humor, and critique. The human pole in H-T-W relations expands from introspection to relational and public selfhood (cf. narrative identity in Loidolt 2018); the technology shifts from interlocutor to mediator of public discourse; and the world opens as a shared horizon of meaning where users negotiate algorithmic understanding. In Arendtian terms, the self gains reality through appearing among others. The loop thus reconfigures subjectivity and publicity: affective intimacy becomes a condition for reflective plurality.

4. *Temporal sedimentation and co-evolution*. Over time, these recursive cycles sediment into habitual expectations. Users learn to prompt the model, anticipate its tone, and manage self-presentation; the algorithm, through retraining and feedback, adapts in turn. The result is not a static relation but co-evolution: the human self acquires new interpretive habits and moral sensibilities, the technology develops new interaction patterns, and the world disclosed through their engagement gains new contours of intelligibility, shaped by emotional resonance, communicative fluency, and algorithmic style. Meaning emerges as an ongoing negotiation among H-T-W poles: open-ended, provisional, and revisable.

Fisher (2022) asks, "Can algorithms tell us who we are?" and answers negatively, warning that algorithmic profiling replaces introspection with computation. While this critique rightly identifies the risk of reducing personhood to data, it remains tied to the dualisms it contests: human versus machine, inner versus outer, authenticity versus simulation. The LLM-biography case unsettles this opposition. In Chandler's blog, CNET's mirror experiment, and the Reddit thread, LLMs do not simply represent the self but elicit self-interpretation. Users judge, correct, and publicly test what the model proposes; their agency is distributed, not eclipsed. Santarpi's therapy sessions suggest similar affective structure, as participants move between accepting the poem's empathy and reclaiming authorship through editing. Against Fisher's binary, algorithms do not tell us who we are; they provide conditions of appearance that users actively negotiate.

Thus, the epistemic model of revelation gives way to a phenomenological model of co-appearance. This also reworks the premises identified in section 1: against the exteriorization of technology, the algorithm is not a "black-boxed other" acting on a passive subject but a constitutive medium within the shared world of appearance and dynamic interpretation. Against the interiorization of politics, meaning arises not in private cognition but in public interpretative exchanges, as in Reddit commentaries or classroom discussions about AI-authored statements. In these micro-publics, judgment is exercised, plurality restored, and the political potential of technological mediation realized.

The practice of generating autobiographical texts with LLMs shows how selfhood unfolds through technological mediation. Across blog posts, online communities, and therapeutic experiments, users enact the hermeneutic lemniscate through affective openings, friction, public re-articulation, and sedimentation. These movements also enact an Arendtian transition from private experience to shared worldliness. The comfort offered by LLMs is therefore not the endpoint but the beginning of political inquiry. When users resist that comfort, correct the AI, or expose its portrayals to public scrutiny, they exercise judgment in Arendt's sense, orienting themselves in a world of plural perspectives. Such hermeneutic appropriation has political force by resisting the quiet normalization of neoliberal meritocracy and homogenized selfhood that LLMs tend to reproduce (Crawford 2016).

The politics here is not antagonistic but *compositional*: it shapes standards of appearance by which social worth is judged. Users' interpretive interventions instantiate Baş's (2026) concept of "technopolitical mediation": technologies mediate the common world of meanings and the intersubjective relations sustaining it. These appropriations blur the boundary between individual and collective judgment. Each LLM-composed biography helps redefine legitimate self-presentation; each public conversation about AI authorship reconfigures shared evaluative frameworks. This shows how postphenomenological hermeneutics, enriched by Arendtian politics, can analyze AI's political dimension without internal transparency. Politics unfolds in the iterative, interpretive co-constitution of appearances—whether in organized resistance or mundane digital self-crafting.

The LLM-biography is thus an interesting laboratory of *political postphenomenology*. It shows how algorithms mediate the appearance of selves and open space for collective reflection on what counts as a self. Algorithms do not abolish the public sphere; they reorganize its texture, transforming spectatorship into co-spectatorship, introspection into dialogue, and biography into iterative world-building. The LLM-generated self embodies this paper's central claim: to understand AI's political dimensions, we must attend not to the hidden "black box" but to the relational, temporal, and worldly processes through which humans understand themselves and appear through technologies. The concluding section builds on this insight to articulate a broader political postphenomenology of AI, where judgment and plurality form the heart of technological politics.

## 5. From mediation to political ontogenesis

In this chapter, we argued that AI politics does not hinge on uncovering an interior "truth" of code but on analyzing how technologies configure appearances over time. Read through postphenomenology, this means tracing how technologically mediated experience interpretatively moves from affective uptake to reflective engagement and public articulation, allowing private resonances to become communicable and "politically real." The biographic LLM case shows how algorithmic amplification makes certain traits of self and world salient for judgment by others, while reduction may background others, shaping the horizon of judgment (Ihde 1990; Verbeek 2005). Rather than treating amplification as reinforcement and reduction as loss, we located their political import in how they compose conditions of appearance through hermeneutic (Kudina 2023) and political (Longo 2025; Baş 2026) postphenomenology. AI's political significance then lies in mediation's temporal world-making: moving users between intimacy and exposure, resonance and contestation, validation and critique.

The paper's methodological upshot is a relational–temporal analytic that couples Arendt's account of spectatorship and judgment with postphenomenological mapping of interpretative dynamics. Practically, this means tracking sense-making co-evolution across H-T-W poles, reconstructing stages of experience without reifying them, and analyzing how amplification/reduction moves from private experience to public evaluation. This approach can be further articulated and refined in application to other cases, e.g., recommender systems, interface nudges, and non-digital technologies by asking how devices compose scenes of appearing, invite salience, background alternatives, and sediment into habits, narratives, and publics. For designers, it suggests the importance and possibility of cultivating *productive friction*: developing interfaces that enable articulation, revision, plurality, contestability, and alternative pathways rather than only comfort and fluency. For policymakers, it urges to look *beyond transparency*: e.g., supporting participatory design and evaluations of AI that surface its lived effects, and alongside audits of models and data, assessing how AI systems make people visible to others and how users appropriate AI.

Our thesis is straightforward: political sense-making does not await transparency but arises within technologically, intersubjectively, and temporally mediated interpretive practices. The “black box” metaphor neglects this dynamic by treating politics as interior and algorithms as external interventions. Here, finally, is our response to the “political deficit” charge: when critics claim that postphenomenology is too individualist, they project onto it the individualist picture of human–technology relations that more mature developments of postphenomenology have precisely rejected. Mediation is not a private interface between a solitary user and a neutral tool; it is a worldly, intersubjective, and temporally unfolding interpretative process in which selves and publics co-appear.

Read with Gadamer and Arendt, this means that the so-called “micro-level” is already saturated with publicity: affective uptake becomes communicable, reflection becomes accountable, and uses sediment into shared standards of appearance. Far from bracketing power, this approach locates it in the conditions under which things and people appear to others, are judged, and reconfigured. Thus, the alleged individualism of postphenomenology is a mirage produced by a static, inside/outside model of technology and a snapshot view of experience. Once mediation is treated as relational and temporal, the political is not an add-on at the “macro” level; it is intrinsic to the very movement through which technologies amplify and reduce, stabilize and unsettle, and thereby compose a shared world in which judgment can take place.